\documentclass[11pt,a4paper,multicol]{article}
\pdfoutput=1
\usepackage{harvard}
\usepackage{graphicx}
\usepackage{graphics}
\usepackage{multicol}
\usepackage{hyperref}
\hypersetup{pdftex,
colorlinks=true,%
linkcolor=red,%
urlcolor=blue}

\newcommand{\be}{\begin{equation}}
\newcommand{\ee}{\end{equation}}
\newcommand{\ba}{\begin{array}}
\newcommand{\ea}{\end{array}}
\newcommand{\bea}{\begin{array}}
\newcommand{\eea}{\end{array}}

\newcommand{\lboro}{
\small{Department of Mathematical Sciences, Loughborough University, U.K.}}

\newcommand{\liege}{
\small Department of Psychology, University of Li\`ege, Belgium}

\begin{document}

\title{Extremely Dilute Modular Neuronal Networks: \\ 
Neocortical Memory Retrieval Dynamics\footnote{\mbox{Pre-peer-review version of Ref.: C. Fulvi Mari, {\it J. Comput. Neurosci.} {\bf 17}: 57--79 (2004). DOI: \href{https://doi.org/10.1023/B:JCNS.0000023871.60959.88}{10.1023/B:JCNS.0000023871.60959.88}}}
\author{Carlo Fulvi Mari\footnote{\href{https://orcid.org/0000-0002-5828-9412}{https://orcid.org/0000-0002-5828-9412}} \\
\lboro \\
\liege \\
\date{24 October 2002}}}

\maketitle

\abstract{A model of the columnar functional organization of neocortical association areas is studied. The neuronal network is composed of many Hebbian autoassociators, or {\it modules}, each of which interacts with a relatively small number of the others. Every module encodes  and stores a number of elementary percepts, or {\it features}. Memory items, or {\it patterns}, are peculiar combinations of features sparsely distributed over the multi-modular network. Any feature stored in any module can be involved in several of the stored patterns; feature-sharing is in fact source of {\it local ambiguities} and, consequently, a potential cause of erroneous memory retrieval activity spreading through the model network. 

The memory retrieval dynamics of the large multi-modular autoassociator is investigated by means of quantitative analysis and numerical simulations. An {\it oscillatory} retrieval process is found to be very efficient in overcoming feature-sharing drawbacks; it requires a mechanism that modulates the robustness of local attractors to noise, and neuronal activity sparseness such that quiescent and active modules are about equally noisy. Correlated activation of interconnected modules and extramodular neuronal contacts more effective than the intramodular ones seem to be general requirements in order to efficiently achieve satisfactory quality of memory retrieval.

It is also shown that, even in ideal conditions, some spots of the network cannot be reached by retrieval activity spread. The locations of these {\it activity isles} depend on the pattern to retrieve and on the cue, while their extension only depends on architecture of the graph and statistics of the stored patterns. The existence of these isles determines an upper-bound to retrieval quality that does not depend on the specific retrieval dynamics adopted, nor on whether feature-sharing is permitted. The oscillatory retrieval process nearly saturates this bound.}\\

\noindent{\bf Keywords:} Modular network; autoassociator; memory retrieval; feature sharing; neocortex architecture.

\begin{multicols}{2}
\section{Introduction} \label{introduction}

The neocortex presents several levels of architectural and functional modularity \cite{Kaas87,Brait,Fuster97,Mount97}. What seem to be the elementary processing units of cognitive functions are the {\it columns}, compact assemblies of densely interconnected excitatory and inhibitory neurons that extend vertically through the layers of the cortical sheet; the width of the columns ranges between 0.3 and 1.0 mm across the cortex quite independently of the size of the brain, which can span several orders of magnitude across mammalian species \cite{Mount2,Brait,Mount97}. Neuropsychology lesion studies cannot investigate the cognitive relevance of single columns, mainly because of their small size. Imaging techniques (e.g., PET and fMRI) also are not yet resolutive enough to monitor the activity of individual columns. However, neurophysiology has provided plentiful evidence of columnar functional individuality. 

In somatosensory cortex SI, neurons belonging to the same column have very similar and most overlapping receptive fields on the skin; on the contrary, the receptive fields of neurons that belong to different columns are markedly distinct \cite{Mount57}. In visual cortex V1, columns respond preferentially to visual stimuli shown in a well-defined region of the visual field of one of the eyes; when tested with short straight line stimuli, they are selectively sensitive to lines at particular angles of orientation \cite{HubelWiesel}. In medio-temporal area MT, columns are selectively activated by motion directions \cite{Albright}, while columns of the nearby area MSTd are responsive to characteristic combinations of the visual flow stimuli contraction, expansion, rotation and translation, i.e. spirals, according to a continuous tuning curve \cite{Graziano}. 

Columns in inferotemporal cortex (IT) respond selectively to non-elementary visual stimuli that range from moderately complex shapes to figures very rich in details like individual faces \cite{Miya88,Fujita,Tanaka}. According to \citeasnoun{Tanaka},  neurons in area TE may be also sensitive to orientation, size, and contrast polarity of their critical visual features, though being neutral to the object position in their large visual fields.  

The posterior parietal cortex (PPC) seems to be responsible for higher perceptual memory and processing \cite{Mishkin,Mount84}; its participation is evident in tasks involving the actions in, or the perception of, or the attention to the environment that surrounds the subject. Although no elementary sensorial input seems to be able to select columnar activity in PPC, studies on complex stimuli and behavioural response reveal that the neurons of PPC that have similar properties are in fact arranged in vertical columns \cite{Mount75,Mount95,Mount97}: a column of PPC can be selectively active during fixation of gaze, or slow pursuit tracking, or reaching by an arm, or manipulation, or complex visual stimulation. 

Rich of axonal connections with subcortical, limbic and neocortical areas, like the reciprocal connections with the posterior parietal and inferotemporal cortices, the prefrontal cortex (PFC) is at the top of the hierarchy of motor memory\footnote{In general, the activation of a motor representation in PFC is more evident when the subject is learning a new motor task. After practise, it seems that the `schemas' are relocated in motor areas that lie in lower levels of the motor hierarchy \cite{Jenkins}.} and also supports high-level cognitive/executive functions other than pure motor planning \cite{MartinChao,Fuster95,GoldRak,Grafton,Buckner,Savage,Fuster97}. During {\it working memory} performance, when subjects are required to keep prolonged memory of briefly presented stimuli in order to execute a {\it delay task} correctly, marked self-sustained neuronal activity is found in columns of PFC (\citeasnoun{Fuster99} and references therein, e.g., \citeasnoun{Funahashi}). In fact, working memory is not peculiar of prefrontal cortex \cite{Fuster98,Fuster97}: parietal areas simultaneously also maintain high activity level since a perceptual stimulus is processed within its environmental context \cite{FriedmanGoldman}, and, if the stimulus involves visual recognition, inferotemporal regions keep high activity too \cite{FusterJervey,Miya88}. In support to the theory of parallel processing in prefrontal and parietal associative areas, there are also some anatomical findings according to which posterior parietal and dorsolateral prefrontal cortices project in common to virtually the same targets in over a dozen distinct cytoarchitectonic areas \cite{Selemon}. It is often hypothesized that the self-sustainment of enhanced activity is provided by dynamical attractor properties of the columnar neuronal networks and is possibly related to the retrieval of memory features previously learnt.  
 
The prefrontal, posterior parietal and inferotemporal cortices are usually referred to as the {\it association} areas of the neocortex and are thought to also subserve semantic memory \cite{MartinChao,Grafton,Ricci,GoelDolan,Fuster97,Shallice88}, that is, general knowledge of objects and events that is not strictly related to, and does not necessarily depend on specific episodic contexts. The present paper is focused onto modelling memory processes that supposedly take place in the neocortical association areas.

Because intracolumnar {\it recurrent} neuronal contacts are Hebbian and relatively dense \cite{Brait,Mount97}, any column may hold autoassociative abilities, like pattern completion and self-sustainment of structured activity. Indeed, several models have shown that, if a pattern of neuronal activity is allowed to produce Hebbian-like modifications of the recurrent synapses, then a stable dynamical attractor is created in which the network will reproduce the pattern: if at any time an external stimulus, or {\it cue}, puts the network within the basin of attraction, that is, vaguely speaking, if it forces the network to produce a pattern of activity someway similar to the one previously stored, then, under appropriate conditions, the network dynamics will drive the neurons to reproduce the activity they had during the presentation of the original pattern (\citeasnoun{Little}, \citeasnoun{Hopfield}, and a vast subsequent literature). The network can store several patterns, creating an equal number of attractors. The existence of an individual attractor corresponding to each stored pattern permits recovering specific distributions of neuronal activity starting from just a fragment of the original ({\it pattern completion}); once a pattern is in this way recovered from memory, the network can stay in the retrieval attractor for prolonged time even in the absence of an external stimulus. The properties of prolonged self-sustained activity and of pattern completion in recurrent Hebbian network models seem compatible with experimental findings of persistent delay activity \cite{FusterJervey,Miya88,Funahashi} and of neuronal responses that reveal ongoing memory retrieval and recognition processes, especially in association areas \cite{Naya,Tomita,Hasegawa,Fujita,Tanaka,Hoesen}. Together with the evidence of feature selectivity of columns, briefly reviewed above, these further observations suggest that memories may consist of distributed representations\footnote{The hypothesis according to which any mnemonic representation is distributed across areas rather than being localized in a specific locus is widely supported by neurophysiological and neuropsychological/imaging evidence (cf., respectively, \citeasnoun{Fuster97} and \citeasnoun{MartinChao} for reviews; also cf. \citeasnoun{Ishai}).} of which columns locally store the finest elements. It is sometimes hypothesized that one of the advantages of a modular structure of this kind might be the possibility of representing new patterns making use of elements that already appeared in patterns previously stored. The converging evidence from several levels of investigation naturally confers great importance to understanding how the multitude of columns/modules may be organized in order to perform cognitive tasks, first among these the retrieval of neuronal representations from memory, like, for example, in pattern completion tasks. 

Some authors have addressed problems concerning memory retrieval in modular network models of cortical areas \cite{OKT,Papik97,FMT,Parga,Parga2,CFMjpa} and provided results that elucidated some properties of modular autoassociators, but several questions are still open and further investigations seem necessary. In particular, it has not been previously investigated the dynamics of cued retrieval in many-columns autoassociators in the case in which elementary memory features that are stored in any module can participate in several complex patterns, in a system of {\it feature-sharing} memory storage. 

In the present paper, dynamical properties of memory retrieval processes in a feature-sharing modular autoassociator are studied. The network model is composed of a large number of interacting modules; every module is a neuronal (Hebbian) autoassociator, intended to model a single cortical column, and is the site of storage of a number of elementary memory features. The appearance of any locally stored {\it feature} in possibly several global {\it patterns} is what creates most of the problems in building a theoretical model able to reproduce proper memory retrieval and that would also comply with biological constraints. For example, let $A$ and $B$ be two modules connected with each other, and assume that $A$ is elicited by the cue stimulus to retrieve feature $a$, while $B$ is quiescent, in a spontaneous, uniform activity state. As $a$ appeared in several of the learnt patterns, $B$ is likely to have stored more than one feature, e.g., $b^{a}_{1}, b^{a}_{2}, \dots, b^{a}_{n}$, that were respectively active in $n$ patterns that simultaneously also produced feature $a$ in module $A$, that is, in $n$ patterns that share feature $a$ in module $A$. Therefore, Hebbian modifications of synapses between neurons of the two modules keep memory of the multiple associations between feature $a$ of module $A$ and each of the corresponding features $b^{a}_{1}, b^{a}_{2}, \dots, b^{a}_{n}$ of module $B$, thus constituting a source of {\it local ambiguity} in the retrieval process. Consequently, the retrieval of feature $a$ in module $A$ would equally tend to drive module $B$ toward any one of the local attractors corresponding to the $n$ features $b^{a}_{1}, b^{a}_{2}, \dots, b^{a}_{n}$ stored in module $B$, not necessarily the correct one for what concerns the global pattern the cued fragment was taken from. The network must then be able to favour the spreading of correct retrieval across modules while suppressing spuriously activated features. One of the main obstacles is that every module `perceives' the state of only a very small fraction of the others; this condition can potentially lead to the nucleation of regions of the network that retrieve wrong memory items. Increasing largely the average number of modules from which any module receives afferents may seem a way to solve the problem; however, because experimental data constrain the number of extracolumnar inputs to any neuron to be about equal to the number of the intracolumnar inputs \cite{Brait}, it is unfeasible to design long-range ({\it white matter}) afferents from a large number of columns, as this would make the signals from any pre-synaptic module largely overcome by local signals. 

Here, a dynamical mechanism is proposed that can surmount the problem of local ambiguity. The retrieval process relies on periodical oscillations of the robustness of local attractors to noise: During any high-robustness semi-period, active modules are stable and can spread retrieval activity to their neighbours according to Hebbian associations; during the following low-robustness semi-period, only those modules whose activity is supported by at least two of their respective neighbours are not destabilized to quiescence. It follows that the retrieval of local features that are appropriate to the cue is greatly advantaged over incorrect activations. This {\it oscillatory retrieval process} can achieve a retrieval performance in fact only slightly inferior to the one that would be achieved if feature sharing was not allowed (and, hence, simpler non-oscillatory dynamics could be adopted). Interestingly, physiological mechanisms that can modulate the robustness of columnar attractors have been recently shown to exist (cf. \citeasnoun{Durstewitz2} and references therein).

The retrieval performance of this kind of modular memory system, however, is here shown to be necessarily below perfect recollection. Indeed, the extreme dilution of the modular connections and the sparseness of the pattern representation, together with the memory cue being only a small fraction of the pattern to retrieve, determine the existence of {\it activity isles}; these are regions of the network that cannot be percolated by the retrieval activity spreading and that are present independently of the specific retrieval dynamics that is adopted, even when feature-sharing is not permitted. The locations of the isles depend on the specific pattern to retrieve and respective cue stimulus; instead, the fraction of the modular network that in any pattern belongs to activity isles depends only on the architecture of the connection graph and on the statistics of the patterns, thus determining a general upper-bound to retrieval quality. The oscillatory retrieval process almost saturates this bound. It should be emphasized that, as the present model does not require topographical organization, the modules belonging to the same activity isle are usually scattered throughout the network. It might be conjectured that the existence of activity isles underlies phenomena of incomplete recollection, e.g., in {\it tip-of-the-tongue} states \cite{Brown91,Koriat93}, though speculations in this direction are not pursued further in this paper.

Because any module interacts with only a relatively small number of the others, it seems generally convenient that interconnected modules store features of correlated kinds. Indeed, correlation of the activity of modules that contact reciprocally is here found to be potentially very relevant to memory retrieval: a network that is able to perform excellent cued retrieval from the memory of a large set of patterns with correlated modular activation can produce only poor retrieval if, instead, connected modules are independently recruited across the set of memory patterns, all the other constructive parameters staying the same. 

To the knowledge of the author, both the study and proposed solution of the problem of local ambiguity and the introduction of the concept of activity isles are novel contributions to the literature of many-columns autoassociators. 

In Section \ref{methods}, the multi-modular network model is defined: neuronal model (\ref{model}), architecture of axonal projections (\ref{architecture}), and correlational scheme (\ref{statistics}) are described. In Section \ref{fieldov}, some statistical quantities are identified concerning neuronal input currents and their correlation with stored features. Neuronal signal-to-noise ratio is calculated in Section \ref{StoN} and used as an indicator of attractor stability. In Section \ref{dynamics}, dynamical transitions of the state of any module are assigned probabilities that are functions of the signal-to-noise ratio and of the state of the modules that contact the former; results from simulations of the dynamical network model are presented and discussed. The simpler case in which local features are not shared among different patterns is illustrated in Section \ref{isles}, where the concept of activity isles is also introduced. Conclusions are drafted in Section \ref{conclusions}.

\section{Methods}\label{methods}

\subsection{Neuronal model}\label{model}
Firing-rate coding is adopted and it is assumed that in any {\it active} module any neuron can be either in an excited or in a suppressed firing level; this approximates the bimodal distribution of firing-rates found in neuronal networks during persistent ({\it delay}) selective activity, when it is usually possible to discriminate between {\it foreground} neurons, that fire spikes at high rate, and {\it background} neurons, that fire at very low rate. In the {\it quiescent} modules, all neurons fire spontaneously at the same firing rate, that is higher than the background, and lower than the foreground neurons' firing-rate. An active module is a module that is reproducing any of the local neuronal representations ({\it features}) previously stored.

The firing rate of neuron $i$ in module $m$ is represented by  
\be
V_{i_{m}}=B \tau_{m}' \eta_{i_{m}}^{d} + A (1-\tau_{m}') + C, 
\label{frate}
\ee
where: $\tau_{m}'$ is equal to 1 if module $m$ is active, that is, if it is retrieving a local feature ($d$), and is equal to 0 if the module is quiescent; $\eta_{i_{m}}^{d}$ is equal to 1 if neuron $i$ of module $m$ is in the foreground of local feature $d$, and is equal to 0 if the neuron is in the background; $B+C$ is the firing-rate of foreground neurons; $A+C$ is the spontaneous activity firing-rate; $C$ is the background neurons' firing-rate. If the multi-modular network is reproducing the whole pattern $p$ correctly, then the expression of the firing-rate of neuron $i_{m}$ becomes
\be
V_{i_{m}}=B \tau_{m}^{p} \eta_{i_{m}}^{d_{m}(p)} + A (1-\tau_{m}^{p}) + C, 
\label{frate2}
\ee
where $\tau^{p}_{m}$ is equal to 1 if module $m$ is actively involved by pattern $p$, that is, if a feature stored in module $m$ takes part in pattern $p$, and is equal to 0 otherwise, and use is made of the mapping  
\be
(m,p) \longrightarrow d_{m}(p),  
\label{mapping}
\ee
that produces the local feature $d$ of module $m$ associated with global pattern $p$ (cf. Section \ref{statistics}). The fraction $a$ of neurons in foreground of any feature is called {\it neuronal (activity) sparseness}, and is of the order of $10^{-2}$, consistently with estimates from experiments on inferotemporal cortex by \citeasnoun{Miyashita}. The neurons forming the foreground of any feature in the model are randomly chosen with probability $a$, independently. The statistics of the binary variables $\{\tau^{p}_{m}\}$ in the set of global patterns is detailed in Section \ref{statistics}. For convenience, the active and the quiescent modules of any stored pattern $p$ will be said to constitute, respectively, the {\it foreground} and the {\it background} of pattern $p$, analogously to the definition used for neurons.
 
The parameters $A$, $B$ and $C$ enter the mathematics of the model only through the ratios $A/B$ and $C/B$. The local recurrent circuits are assumed to keep $A/B$ and $C/B$ nearly constant across the modules in spite of the modular input fluctuations. From suitable literature involving electrode recording in prefrontal and inferotemporal cortices, it can be reasonably extrapolated that $A/B \sim 0.1$, with $C/B$ slightly ($\sim 20\%$) larger than $A/B$. In the following, it will be assumed that $A/B=0.1$ and $C/B=0.12$ (values inferred from the plots of \citeasnoun{Funahashi}). 

The total input current to neuron $i$ of module $m$ is written as  
\be
h_{i_{m}}=\sum_{j_{m} \neq i_{m}} J^{S}_{i_{m} j_{m}} V_{j_{m}} + 
\sum_{n \neq m} \sum_{j_{n}} J^{L}_{i_{m} j_{n}} V_{j_{n}}, 
\label{fieldeq}
\ee
where: index $n$ runs over the total number ($M$) of modules except the module ($m$) that contains neuron $i_{m}$; $j_{m}$ runs over all the neurons of module $m$ with the exception of neuron $i_{m}$ (no self-interaction); $j_{n}$ runs 
over all the neurons of module $n$. Non-specific contributions, like fast inhibition, are not reported explicitly. The synaptic weights are determined by Hebbian covariance rule according to the following formulae:
\be
J^{S}_{i_{m} j_{m}} = \frac{b_{i_{m} j_{m}}}{T} \sum_{p} \tau^{p}_{m} 
\left(\frac{\eta^{p}_{i_{m}}}{a} - 1\right) \left(\frac{\eta^{p}_{j_{m}}}{a} - 1\right) 
\label{localhebb}
\ee
for the intramodular ({\it short-range}) contacts, and
\be
J^{L}_{i_{m} j_{n}} = \frac{c_{i_{m} j_{n}}}{\lambda T} 
\sum_{p} \tau^{p}_{m} \tau^{p}_{n}
\left(\frac{\eta^{p}_{i_{m}}}{a} - 1\right) \left(\frac{\eta^{p}_{j_{n}}}{a} - 1\right) 
\label{remotehebb}
\ee 
for the extramodular ({\it long-range}) contacts. Normalization constant $T$ is equal to $L+N-1$, by definition. Index $p$ runs over the $P$ stored  patterns. The quenched random variable $b_{i_{m} j_{m}}$ determines the dilution of the intramodular axonal contacts: its value is 1 with probability $b$, and 0 with probability $1-b$ for any choice of the ordered pair $(i_{m}, j_{m})$ (that is, the probability for a neuron to receive axonal projection from any other neuron of the same module is equal to $b$, and contacts need not be reciprocal). Variable $c_{i_{m} j_{n}}$ takes value 1 if neuron $j_{n}$ in module $n$ projects onto the dendrites of neuron $i_{m}$ of module $m$; otherwise, it is equal to zero. The architecture of the inter-modular connections is included in the set $\{c_{i_{m} j_{n}}\}$ (cf. Section \ref{architecture} for details). Note that, in learning any pattern $p$, synaptic modification happens between neurons belonging to different modules only when the pre-synaptic and the post-synaptic modules are simultaneously involved in pattern $p$ (that is, $\tau_{m}^{p}=\tau_{n}^{p}=1$), which seems biologically plausible.

The expressions for the synaptic weights (Eqs. \ref{localhebb} and \ref{remotehebb}) are a generalization, to modular neuronal networks, of the formula for the weights of uniform networks with sparse neuronal activity \cite{Tso}. In addition, the present model also includes selective inter-modular connectivity (cf. Section \ref{architecture}) and sparseness in the global pattern of activation of the modules (cf. Section \ref{statistics}). 

The prefactor $1/\lambda$ in Eq. \ref{remotehebb} is necessarily given a value greater than one. Indeed, while half of the inputs to any neuron come from neurons of the same module \cite{Brait}, the remaining half of inputs come from a number of other modules (this fact being mathematically reflected into the denominator $\hat{k}$ appearing in the {\it signal} of Section \ref{StoN}); this partition of extramodular afferents would make the signal from any pre-synaptic module significantly weaker than the local signal. Besides, the synapses between neurons that belong to different modules are modified at a slower rate than the local synapses, as the probability for both pre-synaptic and post-synaptic modules to be simultaneously recruited in a global pattern is significantly smaller than the probability for any module to be recruited, of course. Thus, $\lambda$ should be given a value appropriately to alleviate the disadvantages of extramodular synapses in respect to the local ones. Possible interpretations of $\lambda<1$ from a physiological point of view may be that inter-modular synapses are `faster' in learning \cite{FMT}, that the long-range contacts are intrinsically stronger than the short-range ones, and that apical dendrites convey the post-synaptic potentials to the soma more effectively than the basal ones, as white-matter axons make contacts preferably onto the apical dendrites of the pyramidal neurons while the intramodular contacts take place mainly on the basal dendrites \cite{Brait}. Anatomical separations might also allow for functional modulation of the relative effectiveness of local and long-range contacts (this possibility is not exploited in the present work).

It should be noted that any local feature in any module likely appears several ($\sim \nu$) times during learning, the Hebbian local synapses thus being incremented correspondingly; on the contrary, very rarely adjacent modules learn the same pair of features for more than one pattern (cf. Section \ref{statistics}). The reinforcement of local synapses unbalances the contributions of local and 
\end{multicols}

\begin{figure}[t]
\begin{center}
\includegraphics[width=10cm]{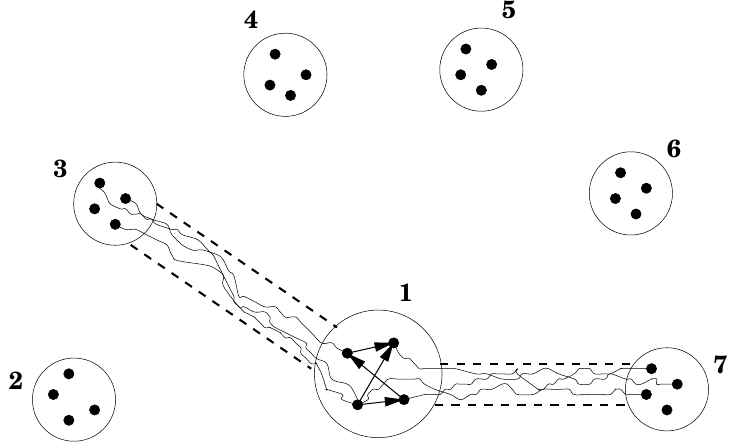}
\caption[]{\footnotesize{Sketch of the structure of connections in the model. 
Black dots represent neurons; circles outline modules. 
The neurons of module 1 connect with each other and receive extramodular 
axonal projections from modules 3 and 7 only. The long-range axons travel through imaginary channels. For clarity, only the intramodular and the extramodular projections relative to the neurons of module $1$ have been drawn. Notice that, although the figure includes small numbers of modules and neurons, the model actually assumes large numbers.}}\label{modfig} 
\end{center}
\end{figure}  

\begin{multicols}{2}\noindent
long-range signals and produces significant fluctuations of the signal-to-noise ratio across the network (cf. Section \ref{StoN}). 

\subsection{Architecture}\label{architecture}
The number of modules ($M$) and that of neurons inside any module ($N$) are 
supposed to be large; this allows for neglecting statistical fluctuations of some quantities that are used in the mathematical model, and to approximate some binomial distributions with Poisson ones. These assumptions do not seem in discordance with reality: the number of neurons constituting any cortical column is estimated to be of the order of $10^4$ \cite{Brait}; being the diameter of the base of any column typically between 0.5 and 1.0 mm \cite{Brait,GoldRak}, and assuming that the `flat' extension of a cortical region is about 0.1 m$^2$, the number of columns per region is of the order of $10^5$.   

Every neuron receives axonal contacts from a number of other neurons (Fig. \ref{modfig}). The probability ($b$) for any neuron to receive a projection from another neuron of the same module is assumed to be equal to 20\% \cite{Brait}. The extramodular afferents to any module are assumed to originate from a small number of the other modules that are here called {\it adjacent} modules or {\it neighbours} of the former (though they may be arbitrarily distant as they are randomly scattered through the entire population); this assumption seems in accordance with results from anatomical labelling experiments (cf. \citeasnoun{Brait}, especially pages 144-145, and \citeasnoun{GoldRak}). One may think of this net of long-range connections imagining the existence of a few {\it channels} that connect any module to other modules; inside these channels, and only through these, the white-matter axons are allowed to pass. The net of channels is modelled as a {\it random graph}, being $s'/(M-1)$ the small probability for any pair of modules to be connected by a channel. Since the number of neighbours per module is distributed around a relatively small value ($s'$) even when the ideal `thermodynamic' limit of infinite number of modules is performed, the graph is said to be {\it extremely dilute}. The probability for any neuron of a generic module $m$ to receive a contact from any neuron of an adjacent module is $L/(\hat{k}_{m} N)$, where $\hat{k}_{m}$ is the number of neighbours of module $m$; so, on average any neuron receives $L$ synapses from extramodular neurons. The presence of $\hat{k}_{m}$ in the denominator of the contact probability evidently implies that the total number of afferent axons from any individual afferent module is smaller in modules that have larger number of neighbours. This is biologically plausible as afferent white-matter axons may have to compete for synaptic space onto the target module. Replacing $\hat{k}_{m}$ with its mean ($s'$) has little effect on the results presented in this paper (once some of the other parameters are appropriately modified). 

The channels, in principle, need not be symmetric, that is, channel $(m,n)$ may be generated randomly and independently from channel $(n,m)$. Nevertheless, for simplicity, the channels are here assumed to be symmetrical. Reality is probably in between full asymmetry and symmetry, or at least this is the impression one has when looking at the results in \citeasnoun{GoldRak} and \citeasnoun{Romanski} obtained by means of the {\it double tracing} technique. Indeed, in the dye pictures of \citeasnoun{Romanski} it seems that reciprocal connections between columns are remarkably more frequent than what one would expect in an extremely dilute directed graph, but also several projections seem to be asymmetrical. To the present knowledge of the author, suitable statistics of inter-columnar connections are not yet available. 

The structure of the net of inter-modular connections is included in the set of coefficients $\{ c_{i_{m} j_{n}} \}$ through the factorization
\be
c_{i_{m} j_{n}} = s_{m n} \cdot g_{i_{m} j_{n}},
\ee
where: $s_{m n}$ is the {\it structure} binary variable, whose value is 1 if a channel exists between modules $m$ and $n$ (it does with probability $s'/(M-1)$), and 0 otherwise; $g_{i_{m} j_{n}}$ is equal to 1 if neuron $j$ of module $n$ is pre-synaptic to neuron $i$ of module $m$, given that module $n$ projects to module $m$ (that is, given that $s_{m n}=1$; thus, $g_{i_{m} j_{n}}=1$ with probability $L/(\hat{k}_{m} N)$ if $s_{m n}=1$, being $\hat{k}_{m}$ the number of neighbours of module $m$), and is equal to 0 otherwise. In the following, the numbers of the extramodular inputs and of the intramodular ones are equal to each other \cite{Brait}; for this assumption, the ratio $\gamma \equiv L/T$ must be equal to $b/(1+b)$.

Possible topographies in the distribution of the modules through the neocortical sheet and of their connections are not accounted for. In fact, as the association cortices are thought to be at the top level of the information processing hierarchy and to cooperate in a parallel manner \cite{Fuster98,Fuster97,GoldRak,FriedmanGoldman,MartinChao,Selemon}, it seems a reasonable approximation to assume non-preferential directions or concentrations (that is, isotropy and homogeneity) of long-range projections between their columns. 

\subsection{Pattern statistics. Correlation}\label{statistics}
Every {\it pattern} stored in the multi-modular network corresponds to a specific distribution of neuronal activity. Accordingly with evidence and hypotheses portrayed in Section \ref{introduction}, any pattern is a peculiar combination of local {\it features} sparsely distributed over the network, each feature being stored in a different column.

In order to create memory patterns suitably distributed, the following ideal construction is adopted: It is assumed that every module can nominally store $D$ features, which the $P$ global patterns are randomly and independently associated to. For any module, as it were, $P$ objects are randomly distributed into $D$ boxes; this distribution is uniform, so the probability for any pattern to fall into any box is $1/D$. $P$ and $D$ are assumed to be large and related to each other by the formula $P=\nu D /\tau$, with $\nu$ finite. This procedure defines the mapping $d_{m}(p)$ of Eq. \ref{mapping}, which is then non-invertible. Only some of the patterns associated to any box actually activate the corresponding feature; as any pattern involves any module with finite probability $\tau$, the probability distribution of the number of patterns eliciting a feature in any module is about Poissonian with mean equal to $\nu$. Hence, any local feature may be involved by several patterns. In any module, some boxes may be not associated with any pattern at all, and some may be associated with patterns that do not activate the corresponding features; eventually, the number of features that are actually stored in any module is about $\tau P (1-e^{-\nu})/\nu$.

One of the basic components of the model is the presence of correlation in the activity of connected modules through the set of stored patterns: {\it Any two adjacent modules are simultaneously active or quiescent with probability higher than chance, while the activities of any pair of non-adjacent modules are nearly independent}. Indeed, since modular activity is sparse and every module interacts with only a few other modules ($\sim s'$), it seems convenient that adjacent modules analyse correlated kinds of features and, consequently, can more often transmit useful information to each other, for instance, for the completion of a retrieval task (cf. also Sections \ref{dynamics} and \ref{isles}, especially Fig. \ref{nocorre}). The correlation of the activities of connected modules is represented by the following table of conditional probabilities \cite{FMT,CFMjpa}:
\be
\begin{array}{l}
{\cal{P}}(\tau^{p}_{m} = 1 | \tau^{p}_{n} = 1, s_{m n} = 1 ) = t_{1}  
\\[.4cm]
{\cal{P}}(\tau^{p}_{m} = 1 | \tau^{p}_{n} = 0, s_{m n} = 1 ) = t_{0}  
\\[.4cm] 
{\cal{P}}(\tau^{p}_{m} = 1 | \tau^{p}_{n} = 1, s_{m n} = 0 ) = \tau \\[.4cm] 
{\cal{P}}(\tau^{p}_{m} = 1 | \tau^{p}_{n} = 0, s_{m n} = 0 ) = \tau  \\
\label{stat}
\end{array}
\ee
where $t_{1} > \tau$ and $t_{0} < \tau$, $\tau$ being the probability for any  module to be active, introduced earlier. So, the activation state of module $m$ does not depend on that of module $n$ if they are not connected by a channel.  
Since the structure variables of the graph are symmetric ($s_{m n}=s_{n m}$), Eqs. \ref{stat} require that $(1 - t_{1}) \cdot \tau = t_{0} \cdot (1 - \tau)$. In fact, the table of Eqs. \ref{stat} has to be seen as the average of the actual distribution, that also takes into account statistical fluctuations across the graph. It should be also emphasized that the probability for any module to be recruited in a pattern is assumed not to depend on the number of its neighbours. (Cf. \citeasnoun{CFMjpa} for more details about the probabilistic scheme.)

Proving the existence of at least one joint probability distribution whose averaged marginals are given by Eqs. \ref{stat} is crucial in order to consider the present statistical model meaningful. Indeed, not every arbitrary set of marginal probabilities over a family of random variables is {\it consistent}: there is not complete freedom in choosing {\it a priori} marginal probabilities on a correlated system. In \citeasnoun{CFMjpa} it is shown that the statistics introduced into the present neural model is consistent if probability $t_{1}$ takes values below an upper-bound that is function of the connectivity $s'$ and of the modular sparseness $\tau$. 

Notice that the correlation between the activation states of adjacent modules does not arise from dynamics of the network; it comes from the statistics of the `natural' input patterns, that are composed by kinds of features that are not independent from each other. The statistical dependence is reflected by the inter-modular connections, as connected modules are assumed to `analyse' features that are someway related to each other. It may be conjectured that, during learning, correlated features tend to be stored in adjacent modules. 

\subsection{The field and its overlap with locally stored features}\label{fieldov}

Due to analogies with physics models, the total input current $h_{i_{m}}$ to neuron $i_{m}$ (Eq. \ref{fieldeq}) is often called {\it field}. It is possible to write $h_{i_{m}}$ as the sum of a few terms (Appendix A). Noise due to memory load appears in the final expression of $h_{i_{m}}$ as a set of Gaussian random variables; the other contributions to $h_{i_{m}}$ are combinations of discrete dynamical variables and quenched random variables. 

To understand what state a module is being `pushed' toward by its neighbours, it is useful to know how much the extramodular axonal inputs correlate with any of the features stored in the module (cf. Section \ref{dynamics}). With this purpose, the {\it field-overlap} order-parameter
\be
Q^{\xi}_{m} \equiv \frac{\lambda \hat{k}_{m}}{\gamma (1-a)^{2} B} \frac{1}{N} 
\sum\limits_{i_{m}} (\eta^{\xi}_{i_{m}}-a) h_{i_{m}}^{ext} 
\ee
is defined as a measure of the similarity between extramodular input currents $h_{i_{m}}^{ext}$ and feature $\xi$ stored in module $m$ ($\hat{k}_{m}$ is the number of modules adjacent to module $m$). 

It is convenient to define the binary variables $\{\varphi_{m}\}$: $\varphi_{m}$ is equal to 1 if module $m$ is correctly either quiescent or retrieving the feature that corresponds to the pattern that should be retrieved by the multi-modular network (conventionally assumed to be pattern \#1), and is equal to 0 otherwise. The overlap between the inputs to the neurons of module $m$ and local feature $\xi$ stored in the same module then is 
\be
\bea{l}
Q^{\xi}_{m} = \tau_{m}^{1} \delta [d_{m}(1)-\xi]
\sum\limits_{n \neq m} \varphi_{n} s_{m n} \tau_{n}^{1} + \\[.8cm]
+ \sum\limits_{n \neq m} \varphi_{n} s_{m n} 
\tau_{n}^{1} \sum\limits_{p\in d_{n}(1), p\neq1} \tau^{p}_{m} \tau^{p}_{n} 
\delta [d_{m}(p)-\xi] + \\[.8cm]
+ \sum\limits_{n \neq m} (1-\varphi_{n}) s_{m n} 
\tau_{n}^{p_{n}} \sum\limits_{p\in d_{n}(p_{n})} \tau^{p}_{m} 
\tau^{p}_{n} \delta [d_{m}(p)-\xi],
\eea
\label{overlap}
\ee
where $\delta [i-j]\equiv \delta_{ij}$ is the Kronecker function for the integer variables $i$ and $j$. The Gaussian terms in $h_{i_{m}}^{ext}$ (cf. Appendix A) do not contribute to $Q^{\xi}_{m}$ due to self-averaging (for $N \rightarrow \infty$). It is useful to the rest of the paper to introduce the following definitions:
\be
\ba{l}
{\hat u} \equiv \sum\limits_{p \in d_{m}(1), p\neq 1} \tau^{p}_{m}, \\[.8cm]
{\hat X} \equiv \sum\limits_{n \neq m} \varphi_{n} s_{m n} \tau_{n}^{1},\\[.8cm]
{\hat v} \equiv \sum\limits_{n \neq m} (1-\varphi_{n}) s_{m n} \tau_{n}^{p_{n}},
\\[.8cm] 
{\hat r} \equiv \sum\limits_{n \neq m} \varphi_{n} s_{m n} \tau_{n}^{1} \sum\limits_{p\in d_{n}(1), p\neq1} \tau^{p}_{m} \tau^{p}_{n}, \\ [.8cm]
{\hat f} \equiv \sum\limits_{n \neq m} (1-\varphi_{n}) s_{m n} \tau_{n}^{p_{n}} 
\sum\limits_{p\in d_{n}(p_{n})} \tau^{p}_{m} \tau^{p}_{n}, 
\ea
\label{define}
\ee
where $p_{n}$ is a pattern that recruits in module $n$ the feature that module $n$ is presently reproducing, while $p \in d_{n}(p_{n})$ indicates any pattern $p$ that elicits in module $n$ the same feature as pattern $p_{n}$. The random variables ${\hat r}$ and ${\hat f}$ come respectively from the last two terms of Eq. \ref{overlap} being summed over the features $\xi \neq d_{m}(1)$ of module $m$. The usefulness of the definitions above will appear in the next Sections; for example, ${\hat X}$ is the number of modules pre-synaptic to module $m$ that are retrieving the correct features (correspondingly to the global pattern \#1 to be retrieved), while $\hat{v}$ is the number of neighbours that are retrieving wrong features. The variables ${\hat r}$ and ${\hat f}$ may have relevance in semantic association errors; for example, ${\hat r}$ relates to the number of features that are elicited in module $m$ by the patterns other than pattern \#1 that also recruit in any of the neighbours of $m$ in the foreground of pattern \#1 the same feature as pattern \#1, reflecting local ambiguities in memory associations. 

\section{Results}

The model proposed here does not lend itself to complete analytical treatment, neither by mean-field techniques nor by methods \`a la \citeasnoun{DerridaGardnerZip}. To evaluate the stability of local retrieval attractors, use is made of a signal-to-noise estimation; then, rules for the dynamics of modular activity are defined. At that point, properties of memory retrieval processes in multi-modular networks are investigated by combining mathematical analysis and numerical simulations. The stability analysis could be made more rigorous, for instance, by extending the approach of \citeasnoun{Parga}. However, here it will be sufficient a qualitative understanding of the effects of noise, like, e.g, the fact that the local retrieval attractors of a module with more supporting neighbours (cf. following) are more robust to noise. A signal-to-noise analysis seems to give enough insight on general properties while permitting to avoid further technical complications and the exploration of large spaces of parameters \cite{Parga}.  

\subsection{Signal-to-noise analysis} \label{StoN} 
 
When the number of stored patterns is no longer negligible with respect to the number of synapses per neuron, back-ground memories can affect the quality of the retrieval, or even make it impossible, by contaminating the neuronal inputs with noisy contributions, whose amplitude is randomly distributed across the network according to a Gaussian density function. The standard deviation ($\sigma$) of this noise comes from the calculation of the input field $h_{i_{m}}$ and is reported in Appendix A. There are in fact non-Gaussian noisy terms besides the memory-load ones, as a rapid inspection of the field in Appendix A reveals (they are given by sums of a relatively small number of binary random variables; lines 7$^{th}$ and 8$^{th}$ of the expression of $h_{i_{m}}$ in Appendix A). However, mainly because of the small value of the sparseness $a$ (but also for the values of other parameters, like, e.g., $\nu$), only a small fraction of the neurons in any active module is significantly affected by these non-Gaussian terms. Hence, the latter are neglected in evaluating the stability of local retrieval attractors. (The non-Gaussian terms follow a binomial distribution through the neurons of the module; the stability of the module has then to be evaluated on the basis of the signal-to-noise ratio, introduced below, by compounding binomial and   \\
\rule{85mm}{.1mm}\rule{.1mm}{5mm}
Gaussian distributions. Calculations have been carried out in worst-case scenario that confirm the suitability of neglecting the non-Gaussian terms.) It should be noted that those terms, being correlated to local features, are indeed the responsible ones for biasing attractor switching (cf. Section \ref{dynamics}). Therefore, in the present approximation, the extramodular neuronal noise may be considered as the sum of two components: an unstructured (memory-load, Gaussian) one, tending to destabilize local retrieval attractors but unable to produce active local states, and a structured one, of negligible effect in destabilizing local attractors but able to bias modular activation. Accordingly, an active module cannot move an active neighbour from its local attractor to another feature retrieval if the activity of the neighbour is stable to the Gaussian noise. It may be noticed that, although found in a different framework and model, this inability of any module to drive an active adjacent module into the attractor of another feature is in fact compatible with the behaviour of the 3-module network of \citeasnoun{Parga} in what the authors called {\it independent phase} regime (thanks to the simpler architecture and memory-storage they adopted, \citeasnoun{Parga} were able to use a more detailed firing-rate neuron model and to pursue the analytical approach more extensively). 

The effect of memory-load noise on the stability of local retrieval attractors can be estimated through a standard {\it signal-to-noise} analysis. For example, consider the case in which the network is reproducing a pattern, possibly with some errors or wrong activation of some modules, and consider a neuron belonging to an active module. One component of the field tends to keep the neuron in the correct firing state, and can thus assume two values according to whether the neuron is in foreground or in background; the difference between these two modal values is called {\it signal}. The remaining component of the input field is noisy and tends to disturb the firing level of the neuron. Assuming that the neuron has essentially to distinguish between being in foreground or in background, one can estimate the stability of the state of the module by looking at the ratio between the signal and the standard deviation of the noise
\end{multicols}
\par{
\be
\frac{\cal{S}}{\cal{N}}= \sqrt{\frac{\gamma}{\alpha a \tau}} \frac{1+\hat{u} +\frac{1}{\lambda \hat{k}} \hat{\Gamma}}{\sqrt{\frac{a (B+C)^2 + 
(1 - a) C^2}{a B^2} (1+\nu) +\frac{t_1}{\lambda^2 \hat{k}} 
\left[ \frac{a (B+C)^2 + (1 - a) C^2}{a B^2} (\hat{\Gamma}+\hat{\Xi})+ 
\frac{(A+C)^2}{a B^2} (\hat{k}-\hat{\Gamma}-\hat{\Xi}) \right] }}, 
\label{StoN1}
\ee
\vspace*{1mm}
}
\pagebreak

\begin{multicols}{2}
\noindent
where: $\hat{k}$ is the number of neighbours of the module that contains the given neuron; $\alpha \equiv P/T$ is the memory-load parameter; $\hat{u}$ is defined in Eqs. \ref{define}; $\hat{\Gamma}$ and $\hat{\Xi}$ are, respectively, the number of {\it supporting}\footnote{Any neighbour of module $m$ is here said to be supporting module $m$ if the two modules are retrieving local features that appeared simultaneously in at least one of the stored patterns and, consequently, support the activity of each other through Hebbian association.} neighbours and the number of the other active neighbours. 

Two considerations can be immediately drawn looking at the ratio above: (1) if active modules and quiescent modules are equally noisy to any post-synaptic one, then the denominator no longer depends on the number of neighbours $\hat{k}$; (2) statistical fluctuations of the number of patterns $1+\hat{u}$ that locally share the same feature can significantly affect the value of the ratio. While the fact in point (1) facilitates the use of threshold mechanisms common through the whole network, the fact in point (2) has just the opposite effect. Indeed, the discrete variable $\hat{u}$ follows a Poisson distribution across the modules, with mean $\nu$; evidently, the local retrieval signals of features that appear in many patterns are significantly larger than those of features that only seldom are recruited in global patterns. As a consequence, if the requirement of having local and long-range signals of similar average magnitude holds, the value of the signal-to-noise ratio is heavily affected by this variability. In view of the oscillatory retrieval process introduced in Section \ref{dynamics}, such variability might make it necessary to adopt large noise levels in order to destabilize all spurious states, but with the further consequence of destabilizing many correct retrieval states and then spoiling the global retrieval quality. There is in fact the likely possibility that the number of modules in wrong but, thanks to \\ 
\rule{85mm}{.1mm}\rule{.1mm}{5mm}
the multiplicity $\hat{u}$, very robust activity is not large enough to support wide spreading of wrong activity, even if such modules cannot be destabilized by the noise: when in module $m$ a wrong feature is activated that is shared among a large number of patterns, the low-robustness step (cf. Section \ref{dynamics}) is likely to be unable to destabilize it; in the next, high-robustness time-step the wrong retrieval in module $m$ is then likely to elicit further wrong activation in its neighbours. The activity of the latter may not necessarily 
be stable enough to survive the next low-robustness step, and this fact may be expected to hamper further spread of wrong feature retrieval. However, studying this possibility in simulations would require storing the whole set of patterns, which is not feasible; in the simulations of the present work, it has then been necessary to consider the case in which all wrong activity is destabilized (cf. Appendix B). From Eq. \ref{StoN1}, it is also clear that $\nu$ cannot be arbitrarily large, as in \citeasnoun{OKT} and \citeasnoun{FMT}.

It may be assumed that the memory of any local feature is someway prevented from being reinforced again and again every time the same feature appears in a global pattern to store. The anatomical separation between the synapses of long-range axons and those of local contacts \cite{Brait} may be relevant to this functional difference. Alternatively, some {\it unlearning} processes \cite{Crick83,Hopfield83,Crick95} may be invoked in order to level off the depth of local attractors. When a mechanism exists that levels the strength of the attractors, the signal-to-noise ratio should be expected to take the form 
\end{multicols}
\par{
\be
\frac{\cal{S}}{\cal{N}}= \sqrt{\frac{\gamma}{\alpha a \tau}} \frac{1 +\frac{1}{\lambda \hat{k}} \hat{\Gamma}}{\sqrt{\frac{a (B+C)^2 + 
(1 - a) C^2}{a B^2} +\frac{t_1}{\lambda^2 \hat{k}} 
\left[ \frac{a (B+C)^2 + (1 - a) C^2}{a B^2} (\hat{\Gamma}+\hat{\Xi})+ 
\frac{(A+C)^2}{a B^2} (\hat{k}-\hat{\Gamma}-\hat{\Xi}) \right] }},
\label{StoN2}
\ee
\vspace*{.1mm}
}
\begin{multicols}{2}
\noindent which, evidently, is not affected by featural multiplicity. Although the problem of excessive reinforcement has been often discussed in the past in many contexts, it is still wide open and it is not aim of the present paper to investigate it further. 

The problem of local variability of robustness would disappear if the synaptic weights were organized in such a way that extramodular signals largely overcame the local ones, but this should be also expected to suppress the local recurrent processes and, thus, should be not considered as a realistic alternative. 

The noise has been assumed to be due to memory load and, hence, its intensity depends on the memory-load parameter $\alpha$, that is proportional to the number \hspace*{-1mm}\rule[-5mm]{.1mm}{5mm}\rule{85mm}{.1mm} $P$ of patterns stored. Tuning of $P$ may then be required, as investigated in \citeasnoun{OKT} and \citeasnoun{FMT}. However, noise may be of other origins, in which case its intensity ($\alpha$) would not necessarily depend on the number of patterns stored in the network.

The simulations carried out in the present work made use of the signal-to-noise ratio in the form of Eq. \ref{StoN1}. However, because simulating the network dynamics with the full set of stored patterns is presently unfeasible, $\hat{u}$ in the signal has been the replaced by $\nu-1$, as if any feature were recruited by the same number ($\nu$) of patterns (cf. Appendix B). 

Once the signal-to-noise ratio is available, since noise is Gaussian, one can easily estimate the fraction of neurons that receive inputs nearer to the respective correct modal value than to the wrong one. If the ratio is too small, this fraction does not differ significantly from the chance level 0.5, while if the ratio is large the fraction approaches 1. In order to evaluate the stability of any active module, it is chosen to establish a threshold $\theta$: if the fraction of stable neurons is not smaller than $\theta$, then the active module is assumed to be in a stable attractor; otherwise, the module decays to quiescence. It is a crude  model of destabilization, but not so much for what concerns the sharpness of the threshold; indeed, in many models of Hebbian neuronal network the transition from stability to instability as a function of the noise level is quite abrupt. The probability for any active module to be unstable will be indicated by $H({\hat k},{\hat u},{\hat \Gamma},{\hat \Xi})$, in the present case just equal to either 0 or 1.

The necessity of $\lambda < 1$ has already been commented in Section \ref{model}. From Eq. \ref{StoN1} it also appears clearly that correlation between adjacent modules is considerably useful (especially for small $\tau$): indeed, if $t_{1}$ was equal to $\tau$, then any foreground module would have, on average, a smaller number of neighbours in foreground, with the consequence of smaller contribution of the extramodular signals to retrieval stability and poorer ability of cue completion (cf. also Sections \ref{dynamics} and \ref{isles}), in spite of the total number of foreground modules being the same (about $\tau M$). 

\subsection{Retrieval Dynamics} \label{dynamics} 
The dynamical transitions between modular states are defined as follows:
\begin{itemize}
\item If a module is quiescent, it can be elicited to retrieve local feature $\xi$ by only an extramodular input field that overlaps with that feature ($Q^{\xi}>0$) and only if the new state would be stable (according to the signal-to-noise ratio; cf. Section \ref{StoN}). If the extramodular field overlaps with one only feature whose retrieval would be stable, then the module will retrieve that feature. If $Q^{\xi}>0$ for several features $\xi$ whose retrieval would be stable, then the module will retrieve the feature with the largest field-overlap among them (in the case of multiplicity of the maximizers, the module will retrieve a feature randomly chosen among the latter).
\item If a module is in an active state, correct or not, it is destabilized 
and moved to quiescence if the signal-to-noise ratio falls below a certain 
threshold $\theta$ (cf. Section \ref{StoN}). 
\end{itemize}

\begin{figure}[t]
\vspace{0cm}
\mbox{\hspace{1.5cm} \includegraphics[width=16cm]{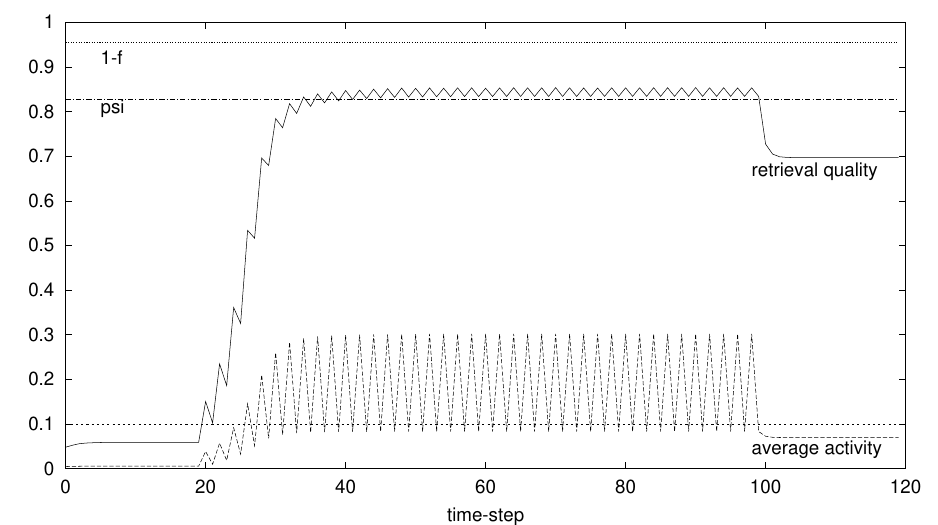}}
\vspace{0cm}
\caption[]{\footnotesize{Temporal evolution of retrieval quality and average activity in cued retrieval simulations. The cue consists of $\sim 5\%$ of the pattern to retrieve. The line ``psi'' reports the present value of $\psi$ (Eq. \ref{psi}), while ``1-f'' is the upper-bound for the case of no feature-sharing (Section \ref{isles}). A line corresponding to the average modular activity in the stored pattern ($\tau$) is also shown for reference. Parameters: $\tau=0.1$; $t_{1}=0.4$; $s'=8$; $b=0.2$; $M=400,000$; $a=0.027$; $\lambda$=0.05; $\nu=10$.}} 
\label{retfig}
\end{figure}

Because of the way the features are distributed across the modules, the probability for any feature in a module to take part in a pattern simultaneously with a given feature in another module is vanishingly small ($\propto 1/D$); consequently, it is very unlikely that the last two sums of the field-overlap (Eq. \ref{overlap}) contribute more than one unit for any feature $\xi$ stored in the post-synaptic module $m$ and different from the correct local feature $d_{m}(1)$, that should be retrieved. As modules in correct active states always cooperate in driving a common post-synaptic module toward the correct state, these observations may suggest that it could be useful to have a {\it field-overlap  threshold} with value equal to 2: if a quiescent foreground module receives a total input field whose overlap with any local feature is not larger than 1, then the module shall not move from quiescence. A simple consequence of this assumption is that the activation of wrong features is mostly inhibited. However, when the cue is only a small fraction of the pattern to retrieve, a field-overlap threshold larger than 1 could also heavily spoil correct retrieval spread; in fact, simulations show that the performance with the field-overlap threshold equal to 2 is very poor already for cues consisting of $\varrho=5\%$ of the pattern to retrieve (cf. following, especially Figure \ref{retfig}, time-steps 0 to 20, and related text). 

If the cue stimulus is small, a low field-overlap threshold would facilitate the spreading of the correct retrieval, but would also allow wrong activation to spread and possibly self-organize in retrieving a pattern different from the one appropriate to the cue. An {\it oscillatory} process is here introduced as a possible solution to this problem: The field-overlap threshold is constantly equal to 1 and the robustness of local attractors to noise periodically oscillates (by modulation of the neuronal threshold $\theta$; cf. Section \ref{StoN}) in such a way that every time-step in which any active module with one only supporting neighbour is stable is followed by a time-step in which it is unstable. In this way, all the modules that have been wrongly activated during a high-robustness time-step are quickly put to quiescence in the next time-step; this is because any module that is in a wrong feature retrieval is most likely supported by no more than one of its neighbours, the same one that drove the module from quiescence to wrong activity. The two-steps oscillation is iterated several times. If any low-robustness time-step was followed by two or more high-robustness time-steps, many of the modules in wrong retrieval would elicit further wrong activity and, consequently, would have more supporting neighbours and gain in robustness. During all the retrieval process, any quiescent foreground module with two or more neighbours in correct active state is driven to the correct local retrieval attractor (if it would be stable). 
\end{multicols}

\begin{figure}[t]
\begin{center}
\includegraphics[width=14cm]{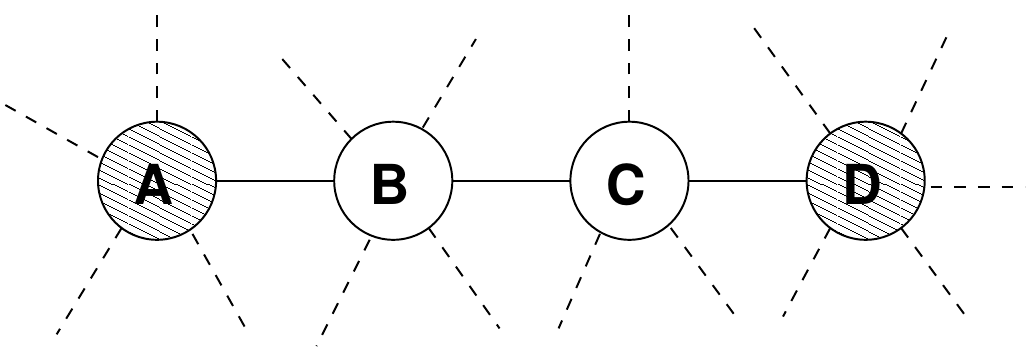}
\caption[]{\footnotesize{Illustration of a simple example of retrieval spreading. Modules $A$ and $D$ are in correct feature retrieval (shaded). Modules $B$ and $C$ are quiescent but should be active in the pattern to retrieve. Cf. Section \ref{dynamics} for details.}} 
\label{spread}
\end{center}
\end{figure}  

\begin{multicols}{2}
A sample case of correct retrieval spread is pictured in Figure \ref{spread}: Modules $A$ and $D$ are in correct feature retrieval, while modules $B$ and $C$ are quiescent but should be active. During a high-robustness time-step, modules $A$ and $D$ elicit retrieval of features in modules $B$ and $C$ (for clarity, the influence of other neighbours to $B$ and $C$ is neglected). If any of the two features activated respectively in $B$ and $C$ is not correct, each of them only has support from one neighbour, that is, respectively, $A$ or $D$. In this case, during the following time-step both modules $B$ and $C$ are destabilized and decay to quiescence. On the contrary, if the activity elicited in $B$ and $C$ corresponds to correct feature retrieval, modules $B$ and $C$ will support each other and still hold the support of, respectively, module $A$ and module $D$. Thus, in the following low-robustness  time-step both of them will be supported by two neighbours and, consequently, will not be destabilized.

Although to date (to the knowledge of the author) there is no direct experimental evidence about such kind of oscillatory retrieval process (though literature on brain waves is abundant), there is already evidence of neuromodulatory mechanisms that could regulate the robustness of local (columnar) attractors to extramodular disturbing signals (\citeasnoun{Durstewitz2} and references therein). It could be noticed that, starting with a cue consisting of about $5\%$ of the pattern to retrieve, about 10 cycles are sufficient to reach plateau-level of retrieval quality (Figure \ref{retfig}: time-steps from \#20 to \#40); at oscillation frequencies of about 40 Hz (in the range of the brain $\gamma$-waves, associated to performance of cognitive tasks), the process would require about 250 ms, a plausible time scale.

The probability for any foreground module to have two or more neighbours also in foreground in any pattern is 
\be
\psi \simeq 1-(1+s' t_{1}) e^{-s' t_{1}}
\label{psi}
\ee
(strict equality holds in the limit $M\rightarrow\infty$); therefore, in static conditions with constant low robustness the retrieval quality cannot be larger than $\psi$. To increase this upper-bound, one can set $t_{1}$ and $s'$ so that $s' t_{1}$ is as large as possible, convening with the correlation bound (as already mentioned, correlation cannot be arbitrarily large, and its upper-bound monotonically decreases as $s'$ increases \cite{CFMjpa}). The modular sparseness $\tau$ also plays a role in this, as larger $\tau$ allows for larger $t_{1}$ given $s'$, within certain limits and biological plausibility. In this view, a convenient choice seems to be $\tau=0.1$ and $s'=8$, that allows for $t_{1}$ as large as 0.4, at least.  

All the possible state transitions of any module are reported in Table \ref{transprob} together with the respective probabilities. Factor $\eta$ takes value 0 or 1 if the field-overlap threshold is equal, respectively, to 2 or 1. The following abbreviations have been introduced: $w$ stands for `wrong', $c$ stands for `correct', $a$ stands for `active', $\bar{a}$ for `non-active', and the upper index $1$ or $0$ indicates, when necessary, whether the considered module is respectively in the foreground or in the background of the global pattern that should be retrieved. For example, $ca \rightarrow w\bar{a}$ indicates the transition of a foreground module from the correct active state to wrong quiescence, while $w\bar{a}\rightarrow wa^{1}$ indicates a transition from  wrong quiescence to a wrong active state. The dynamics is implemented in numerical simulations, with parallel update (some details on the simulations are reported in Appendix B).

In order for the oscillatory process to work properly, it is necessary that active modules are nearly as noisy as the quiescent ones. This condition is realized if
\be 
a = \frac{A}{B} \cdot \frac{A/B+ 2 C/B}{1 + 2 C/B},
\ee
which, with the values of $A/B$ and $C/B$ here adopted, implies $a\simeq0.027$, well compatible with the experimental evidence.
\end{multicols}
\begin{table}[t]
\label{table}
\begin{center}
\begin{tabular}{|c|l|} 
\hline 
& \\
  TRANSITION &  \qquad \qquad PROBABILITY \\
& \\
\hline 
& \\
$ca \rightarrow w\bar{a}$ &  $H({\hat k}, {\hat u}, \hat{X}+\hat{v}, 0)$ \\  \hline & \\ 
$wa^{1} \rightarrow w\bar{a}$ & $H({\hat k}, {\hat u}, 1, \hat{X}+\hat{v}-1)$ \\ \hline & \\
$w\bar{a} \rightarrow ca$ & $\left[ \Theta(\hat{X}-2) + \eta \delta_{\hat{X},1} \frac{1}{{\hat r}+{\hat f} +1} \right] \left[ 1-H({\hat k}, {\hat u}, \hat{X}, \hat{v})\right]$ \\ 
\hline & \\
$w\bar{a} \rightarrow wa^{1}$ & $\eta \left[\delta_{\hat{X},0} \Theta({\hat r}+{\hat f} -1) + \delta_{\hat{X},1} \frac{{\hat r}+{\hat f}}{{\hat r}+{\hat f} +1}\right] \left[ 1-H({\hat k}, {\hat u}, 1, \hat{X}+\hat{v}-1)\right]$ \\ 
\hline & \\ 
$c\bar{a} \rightarrow wa^{0}$ & $\eta \Theta({\hat r}+{\hat f} -1) \left[ 1-H({\hat k}, {\hat u}, 1, \hat{X}+\hat{v}-1)\right]$ \\ 
\hline  & \\
$wa^{0} \rightarrow c\bar{a} $ &  $H({\hat k}, {\hat u}, 1, \hat{X}+\hat{v}-1)$\\ 
\hline
\end{tabular}
\end{center}
\caption[]{\footnotesize{Transitions of modular state and respective probabilities, conditional to the states of the neighbours through the variables ${\hat X}$, ${\hat v}$, ${\hat r}$, ${\hat f}$. The Heaviside step-function $\Theta$ is here defined to satisfy $\Theta(0)=1$; $\delta_{i,j}$ is the Kronecker symbol for the integer variables $i$ and $j$.}}
\label{transprob}
\end{table}
\begin{multicols}{2}
To appropriately destabilize spurious activation (cf. Section \ref{StoN}), the parameter $\alpha$ must belong to a certain range of values. A significant hindrance is the large variability of the signal (Eq. \ref{StoN1}), due to the fluctuation of the number of neighbours across the set of modules (this is a consequence of having adopted a Bernoulli random graph as model of the modular net\footnote{Adopting a different type of architecture would also require proving consistency of marginal distributions and studying how to produce patterns with the correlated statistics in the new kind of graphs.}). Indeed, for example, a noise level that destabilizes the retrieval activity of a module that has one supporting neighbour and a total of 4 neighbours also destabilizes the retrieval activity of a module that has 2 supporting neighbours and a total of 8 or more neighbours. In order to reduce this noxious effect, when testing the oscillatory retrieval process it is chosen to make modules with less than 5 neighbours {\it passive}, that is, they are kept quiescent regardless of the cue and of the inputs they receive from the neighbouring modules; as $s'=8$, the passive modules account for slightly less than 10\% of the network, that, with $\tau=0.1$, would code for about 1\% of the pattern representation. It is possible that the neuronal model adopted here is too crude to counterbalance the fluctuations of the signal-to-noise ratio by other means, but it also seems biologically plausible that the number of columns afferent to any other column should not be too small.

\begin{figure}[t]
\vspace{0cm}
\mbox{\hspace{1.5cm} \includegraphics[width=16cm]{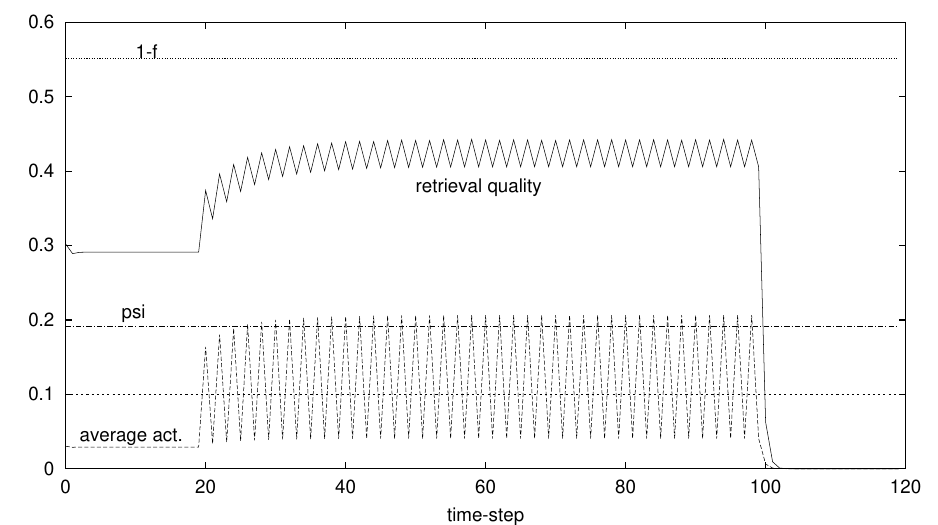}}
\vspace{0cm}
\caption[]{\footnotesize{Temporal evolution of retrieval quality and average activity in cued retrieval simulations in the case in which the activation of adjacent modules is not correlated (Section \ref{dynamics}). The cue consists of about 30\% of the pattern to retrieve. The line ``psi'' reports the present value of $\psi$ (Eq. \ref{psi}), while ``1-f'' is the upper-bound for the case of no feature-sharing (Section \ref{isles}). A line corresponding to the average modular activity in the stored pattern ($\tau$) is also shown for reference. Parameters: $\tau=t_{1}=0.1$; $s'=8$; $b=0.2$; $M=400,000$; $a=0.027$; $\lambda$=0.05; $\nu=10$.}} 
\label{nocorre}
\end{figure}

As explained later in this Section, some spots of the network will not be percolated by correct retrieval activity; the problems mentioned above about handling noise-level also worsen the retrieval capability a little. However, overall performance is remarkably improved by the oscillatory modulation. This is evident in Figure \ref{retfig}, where ``retrieval quality'' is the overlap between the features reproduced by the foreground modules and the pattern to retrieve, while ``average activity'' is the fraction of modules that are active in the network. The cue consists of $\varrho = 5\%$ of the pattern that should be retrieved\footnote{As almost 10\% of the modules are passive, effectively the cue is about 4.5\% of the pattern.} (smaller cues are equally suitable, to some extent, at the expense of longer oscillatory stage). During the first 20 time-steps, the field-overlap threshold is equal to 2 and robustness to noise is high; very little retrieval from memory is produced. This first stage does not belong to the proposed retrieval process and has been included in the simulation just to show how poor the performance would be if the field-overlap threshold was equal to 2 (removing this stage does not affect the other results). At time-step \#20, the field-overlap threshold is lowered to 1 and the oscillatory stage begins; the retrieval quality rapidly increases. (When the cue is very small, e.g. $\varrho=0.01$, the duration of the oscillatory stage necessary to reach the plateau can vary significantly, depending on the random choice of modules that are cued; this variability is due to finite size and becomes soon negligible when larger cues are provided, as for $\varrho=0.05$.) From time-step \#100 onward, the robustness to noise is constantly at the lower level: the retrieved information is stable, there is no spurious activity left (e.g., the average activity is only due to the activity of the modules in correct retrieval, that is,
\end{multicols}
\begin{multicols}{2}
\noindent 
``average activity'' is equal to $\tau$ times ``retrieval quality''), and no further dynamical transition happens. This latest stage is not necessary in the present model and is only included to show the possibility of self-sustainment of retrieval activity even when robustness is low. Stability at this stage is due, of course, to the existence of closed paths of foreground modules through the random graph of connection channels. It should be emphasized that the correlation of modular activation plays an important role for the existence of these closed paths; without correlation, closed paths of foreground modules are too few to permit appreciable self-sustained activity after the oscillatory stage (cf. following, especially Fig. \ref{nocorre}).

If the first two stages of Fig. \ref{retfig} were removed and the cue was provided to the network in the conditions of the third stage, then no significant self-sustained activity would remain because the cue activates modules randomly chosen through the foreground population and has little chance of activating entire closed paths (unless, of course, a much larger cue was provided).

Given any foreground module in correct state, if one or more of its neighbours become active, then the given module always gains in robustness, regardless of the correctness of the new active states of the neighbours: if the activation of neighbour $n$ of module $m$ has been elicited by module $m$, then the feature elicited in $n$ is necessarily one of those that in at least one of the stored patterns were associated with the feature presently retrieved in module $m$, and hence it is supportive to the latter; if, instead, module $n$ is also neighbour to another active foreground module beside $m$, then the present dynamics drives module $n$ to retrieve the correct feature, which of course is consistent with the feature in $m$ and supports it (consider that, if a foreground module is active in the low-robustness step, then it is necessarily in its correct state). As a consequence, the retrieval quality (Fig. \ref{retfig}) can overcome the upper-bound $\psi$ (Eq. \ref{psi}) during the oscillatory phase (arriving not too far, in fact, from the optimal upper-bound $1-f$ calculated in Section \ref{isles}). The final retrieval quality is lower than $\psi$ mainly because of the freezing of modules with less than 5 neighbours and because of the variability of the signal-to-noise ratio as a function of the number of neighbours; minor further corrections arise from the existence of small regions that cannot be percolated by the oscillatory process and, possibly, of modules that do not belong to closed paths of the foreground. 

Simulations have been also carried out in which the noise of neuronal/synaptic origins is neglected and the dynamics `artificially' follows the rule that every module with one only supporting neighbour is unstable during the low-robustness time-steps, while it is stable during the high-robustness time-steps, regardless of the signal-to-noise ratio: the retrieval quality during the
\end{multicols}
\begin{figure}[t]
\vspace{0cm}
\mbox{\hspace{1.5cm} \includegraphics[width=16cm]{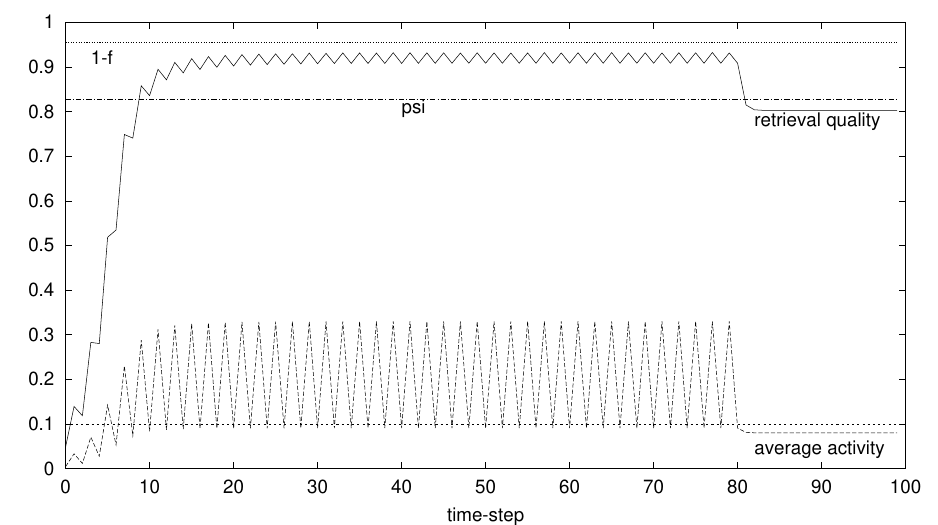}}
\vspace{0cm}
\caption[]{\footnotesize{Temporal evolution of retrieval quality and average activity in cued retrieval simulations with `artificial' dynamics (Section \ref{dynamics}) and no modular freezing. The cue consists of 5\% of the pattern to retrieve. The line ``psi'' reports the present value of $\psi$ (Eq. \ref{psi}), while ``1-f'' is the upper-bound for the case of no feature-sharing (Section \ref{isles}). A line corresponding to the average modular activity in the stored pattern ($\tau$) is also shown for reference. Parameters: $\tau=0.1$; $t_{1}=0.4$; $s'=8$; $M=400,000$; $\nu=10$.}} 
\label{idealfig}
\end{figure}    
\begin{multicols}{2}\noindent
oscillatory stage is only very slightly improved by this proviso, while the final value is significantly larger (+0.05, still below the upper-bound $\psi$; not shown). Of course, removal of modular freezing, no longer necessary in the artificial case, provides significant improvement in both the stages of the retrieval process, as Figure \ref{idealfig} shows (oscillatory stage since the beginning and until time-step \#80). Even in this case, the network cannot recollect the entire memory pattern, because of the existence of subsets of modules that cannot be percolated by the retrieval spread. For example, consider three foreground modules, $A$, $B$, and $C$, such that $A$ is connected to $B$, $B$ is connected to $C$ ($A$--$B$--$C$), and the only neighbours of $B$ in foreground are $A$ and $C$, while the only neighbour of $C$ in foreground is $B$. Suppose also that neither $B$ nor $C$ are elicited by the cue, while the cue or the retrieval process drive $A$ to retrieve the correct feature. During a high-robustness step, $A$ can elicit correct activity in $B$, but in the following low-robustness step $B$ will necessarily decay to quiescence. Hence, $C$ can never be reached by retrieval spread. There are possibly other occurrences of this kind in which retrieval spread is hampered, but the majority of unpercolable modules belong in fact to {\it activity isles}, discussed in Section \ref{isles}. 

Figure \ref{nocorre} shows the output of the simulation of a network identical to the one of Fig. \ref{retfig} but for the absence of correlation in the activity of adjacent modules through the set of stored patterns (that is, $t_{1}=\tau$; stability threshold $\theta$ had to be modified a little, tough the same aim of this change would be achieved by appropriately scaling the noise level $\alpha$ instead). The system initially relaxes to a stable configuration, where it spends the rest of the first stage (removing this stage does not affect the overall performance significantly). When the oscillatory stage begins, there is appreciable memory retrieval, but evidently the performance is very poor (notice the scale of the ordinates), in spite of the cue being six times larger than the one used in the simulation of Fig. \ref{retfig} ($\varrho\simeq 30\%$ vs. $\varrho\simeq 5\%$). This inability should have been expected noticing that the fraction $\psi$ (Eq. \ref{psi}) of modules that have no less than two of their neighbours in foreground is much smaller in the case of uncorrelated activation than in the case of correlated activation (19\% vs. 83\%), even if the sparseness ($\tau=0.1$) is the same. Furthermore, the subgraph constituted by the foreground modules and the channels between them is a random graph of $\tau M$ modules with mean number of neighbours $s' \tau=0.8<1$; this implies that almost all of these modules belong to small, isolated trees \cite{Bollobas}. The instability in the final stage confirms the consequent lack of an appreciable number of foreground modules on closed paths. Instead, in the case of correlated activation the modules of the foreground subgraph have an average number of neighbours about equal to $s' t_{1}=3.2$; therefore, only a 
\end{multicols}

\begin{multicols}{2}\noindent
small minority of modules belong to trees and, in fact, most of the other modules that have at least two neighbours in the subgraph belong to closed paths. These observations suggest that in any case the neocortical columnar networks should have $s' t_{1}$ significantly larger than unity (in the case of uncorrelated activation, $t_{1}$ is equal to $\tau$), at least as long as the architecture adopted here is a reliable model.  

Even when the cue is a very large fraction of the pattern to retrieve, the network of Fig. \ref{nocorre} is not able to produce significant retrieval and to self-sustain activity in the final stage (not shown).   

It may be argued that appropriately increasing the value of $s' \tau$ (e.g., $s' \tau \rightarrow 3.2$) in networks with uncorrelated modular activation could provide them with the same capabilities of the networks that exploit correlated activation. Generally speaking, however, increasing $\tau$ would produce undesirable consequences because the number of features that can be stored in any module cannot be augmented: if the maximum number of stored patterns has to stay the same, then any increase of $\tau$ would imply a roughly proportional increase of $\nu$, which would slow down the retrieval process by increasing (on average) the biasing term $\hat{r}$ (cf. Table \ref{transprob}), beside worsening the possible reinforcement problems commented earlier (cf. Section \ref{StoN}); if $\nu$ should not change, then setting $\tau=0.4$ would decrease the storage capacity four-fold. On the other hand, if the total volume of white matter cannot change, any significant increase of $s'$ would cause, on average, a significant reduction of the effectiveness of the extramodular signals in respect to the intramodular ones, at least at the beginning of the cued retrieval task (cf. Section \ref{introduction}); this may possibly require a decrease of $\lambda$ to an implausible extent and implausible fine-tuning of noise intensity \cite{FMT}. Larger $s'$ also implies larger variance of the number of neighbours across the net, which may affect the efficiency of the threshold mechanisms. From the point of view of biological realism, it could be observed that the larger $s'$, the larger the number of modules that become active during the low-robustness time-steps; for example, with $s'=20$ and $\tau=t_{1}=0.16$, more than 80\% of the network becomes active during these steps, which may be unrealistic. In addition, experimental data seem to suggest that $s'$ is not larger than 10 \cite{Brait}. Finally, for any given $s'$, including also correlation improves the performance, though for large $s'$ correlation may have to be relatively small \cite{CFMjpa}. Therefore, although the values adopted for the parameters of the networks of Figures \ref{retfig} and \ref{idealfig} may be not the optimal ones overall, exploiting the correlation of modular activation through the set of patterns seems to be biologically a convenient solution.
\end{multicols}

\begin{figure}[t]
\begin{center}
\includegraphics[width=11cm]{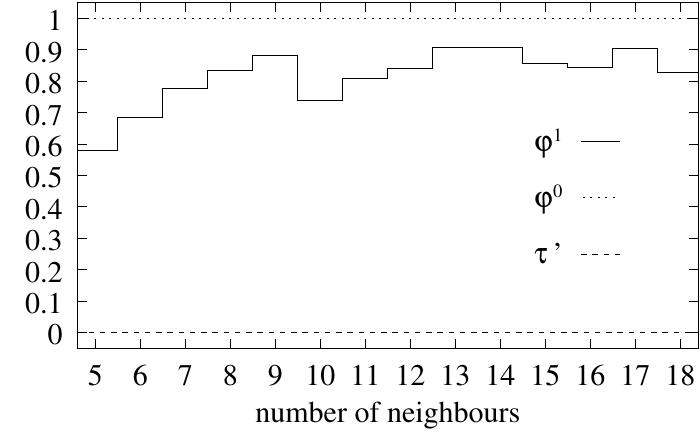}
\caption[]{\footnotesize{Probability for any foreground or background module to be in the correct state (indicated, respectively, by $\varphi^{1}$ and 
$\varphi^{0}$) and for any foreground module that is in a wrong 
state to be active (indicated by $\tau'$) at the end of the simulation of Figure \ref{retfig}, as a function of the number of neighbours. (Modules with less than 5 neighbours are frozen to quiescence.)}} 
\label{histo}
\end{center}
\end{figure} 

\begin{multicols}{2}
Figure \ref{histo} shows the fraction of modules in correct state at the end of the retrieval process of Figure \ref{retfig}, conditionally to the number of neighbours $\hat k$. Variables $\varphi^1$ and $\varphi^0$ are the fractions of, respectively, foreground and background modules that are in correct state, while $\tau'$ is the fraction of foreground modules that are in wrong feature retrieval. The non-monotonicity of the histogram $\varphi^{1}$ is due to the dependence of the signal-to-noise ratio on $\hat k$ (except toward the right end of the plot, where in general statistical fluctuations can be significant). The histograms of $\varphi^{0}$ and $\tau'$ confirm that there is not spurious activity remaining.

When $\varrho$ is large, for example 50\% or more, then the monotonic process with field-overlap threshold equal to 2 and low noise level can achieve larger retrieval quality than the final stage of Figure \ref{retfig} (though not larger than its value during the oscillatory stage), as cued modules that have each only one supporting neighbour do not decay to quiescence. 

\subsection{No feature-sharing. Activity isles} \label{isles}
\begin{figure}[t]
\vspace{0cm}
\mbox{\hspace{1.5cm} \includegraphics[width=16cm]{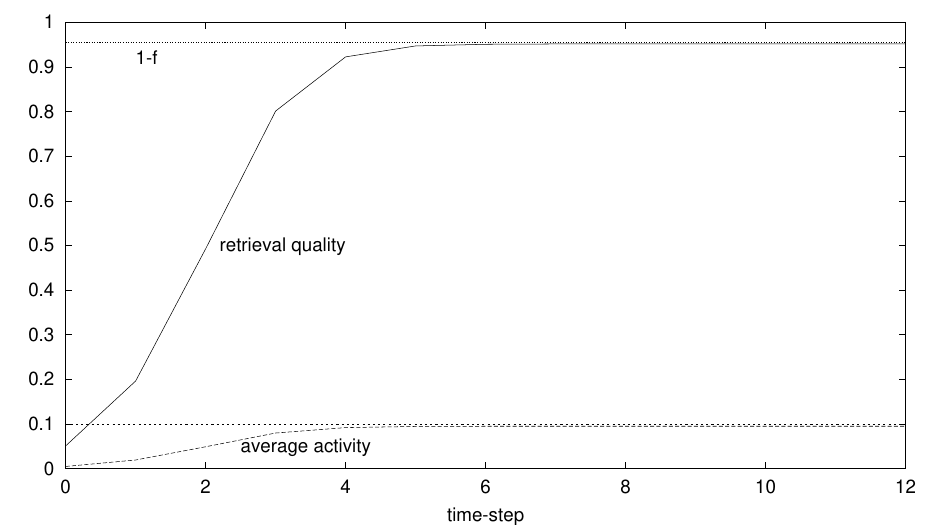}}
\vspace{0cm}
\caption[]{\footnotesize{Temporal evolution of retrieval quality and average 
activity in cued retrieval simulations in the case in which every stored feature appears in only one pattern. The cue consists of 5\% of the pattern to retrieve. The line ``1-f'' gives the maximum retrieval quality permitted by the model (cf. Section \ref{isles}). A line corresponding to the average modular activity in the stored pattern ($\tau$) is also shown for reference. Parameters: $\tau=0.1$; $t_{1}=0.4$; $s'=8$; $M=400,000$; $a=0.027$.}}
\label{noshare}
\end{figure}

\begin{figure}[t]
\begin{center}
\includegraphics[width=8cm]{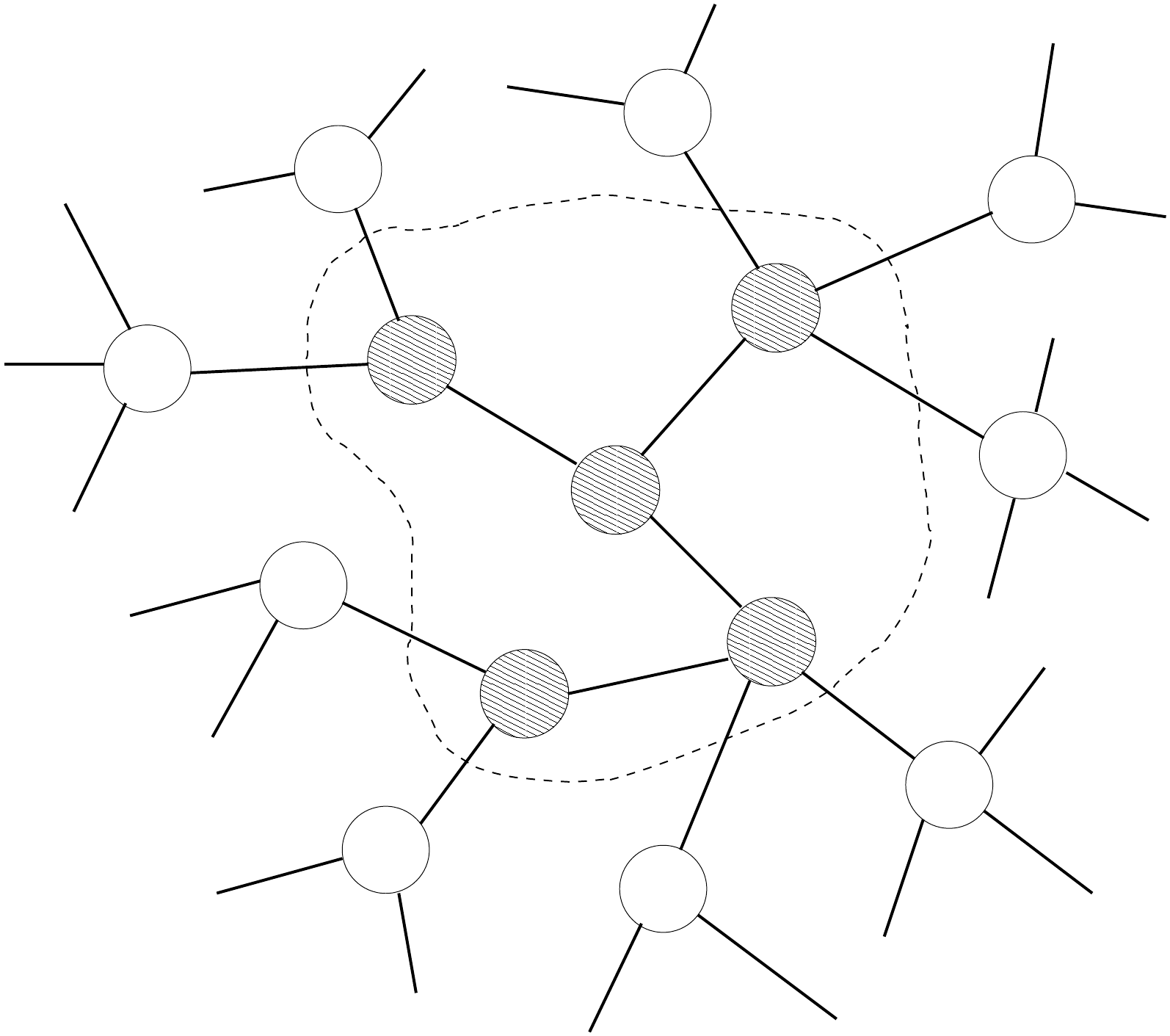} 
\caption[]{\footnotesize{An example of activity isle in the modular network. Shaded circles and empty circles represent, respectively, foreground modules and background modules. The shaded modules constitute an activity isle (dotted line) because the only connections these modules have with the rest of the network are with background modules. (The other neighbours of the background modules are not shown.)}}
\label{exisle}
\end{center}
\end{figure}

As clear from the previous Sections, the main obstacle to obtain proper memory retrieval in the large modular autoassociator is given by feature-sharing: all the wrong activity generated after a `clean' cue is due to the fact that any feature in any module is most likely shared by several of the stored patterns. For comparison, Figure \ref{noshare} shows the outcome of a simulation of a network in which every local feature is active in only one global pattern (so that the mapping of Eq. \ref{mapping} becomes invertible and $\hat{r}=0$); the dynamics adopted here is not oscillatory, as there is no possibility of spurious activations, and no freezing of modules is necessary. As the plot shows, the overlap of the activity with the stored pattern rapidly reaches a value slightly above 95\%. In random graph theory \cite{ErdosRenyi60,Bollobas} it is well established that for $s' > 1$ a finite fraction of the number of modules are directly or indirectly connected with each other, forming the so-called {\it giant component}; the rest form a large number of small, isolated groups. The giant component in a large graph with $s'=8$ accounts for about 99.96\% of the network; thus, the reason of the incomplete retrieval is not the presence of isolated groups of modules. The reason for the phenomenon is in fact that in any stored pattern a certain number of active modules are surrounded by modules that are not active. Given any pattern, let any connected\footnote{A subset of modules is here called {\it connected} if within the corresponding subgraph there exists at least one path that joins any two of the modules.} group of foreground modules that are surrounded by only background modules be called {\it activity isle} (Fig. \ref{exisle}). Many of the isles are not elicited by the small cue, and thus cannot be driven to correct retrieval activity. Consider any $n$ modules in the case in which (1) they form a connected group, and (2) all of them are in the foreground of the pattern to retrieve but none of them is elicited by the cue, and (3) all their neighbours but those in the group are in the background of the pattern to  retrieve; the probability that all these occurrences happen in any group of $n$ modules, that is, the probability that any group of $n$ modules is a non-cued activity isle, has been calculated (the only case of relevance can be shown to be $n \ll M$; details of the calculations are not reported here), from which it follows that the fraction of foreground modules 
\end{multicols}

\begin{multicols}{2}\noindent
that belong to non-cued isles of $n$ modules is
\be
f_{n} \simeq \frac{n^{n-1}}{n!} (s' t_{1})^{n-1} (1 - \varrho)^{n} e^{-n s' t_{1}}
\ee
(strict equality holds in the limit $M\rightarrow\infty$). With the parameters adopted in this case, the value of $f_{n}$ is small and decreases rapidly as $n$ grows: $f_{1}=0.038724$, $f_{2}=0.004799$, $f_{3}=0.000892$, $f_{4}=0.000197$. The fraction $f$ of foreground modules that belong to non-cued isles of any size is the sum of $f_{n}$ over $n$. In fact, simulations show (cf. Figure \ref{noshare}) that the retrieval quality tends asymptotically to the value $1-f \simeq 0.9553$ (but possibly for small statistical fluctuations due to finite size), corresponding to correct local retrieval in all foreground modules but those in the activity isles that are not elicited by the cue and stay quiescent. The existence of activity isles, evidently, does not depend on the specific retrieval dynamics, nor on the adoption of feature-sharing storage. The extent of the network belonging to isles does not depend (in the limit $M \rightarrow \infty$) on the specific cue stimulus and pattern to retrieve. The value $1-f$ constitutes, then, a general upper-bound to retrieval quality. Systems with the same statistics of connections and patterns can at best saturate this bound, regardless of the specific dynamics each one implements.

\section{Discussion} \label{conclusions}

Anatomical and physiological data suggest that association areas of the neocortex could be modelled as a large number of autoassociators (the columns), each of which interacts with a small fraction of the others (Sections \ref{introduction} and \ref{architecture}). If the global memory patterns of neuronal activity are composed of peculiar combinations of local features, each feature being stored in a different module and shared by several patterns, then it becomes important to understand how the multitude of modules is architecturally and functionally organized in order to be able to perform proper memory retrieval when a cue is briefly presented to the network as an external stimulus. One of the major difficulties to overcome in the modelling is that any feature stored in any module is usually involved in several patterns and is thus associated, in a Hebbian way, with several features stored in any of the modules that are in synaptic contact with the former. This implies that, when feature $d_{m}$ is retrieved in module $m$, the latter could move a neighbouring module $n$ to the local attractor of one of the features stored in $n$ and associated with $d_{m}$ during learning which is not necessarily the correct feature for what concerns the retrieval of the pattern corresponding to the cue. 

Investigating retrieval mechanisms that could achieve good retrieval quality without producing erroneous activity, an oscillatory retrieval process is found 
\end{multicols}

\begin{multicols}{2}\noindent
to be particularly efficient (Section \ref{dynamics}). It requires: a modulatory mechanism that periodically modifies the robustness of local attractors to noise; neuronal activity sparseness such that active modules are nearly as noisy as the quiescent ones; number of neighbours that fluctuates from one module to another across the network significantly less than the degree of the nodes in Bernoulli random graphs. In order to efficiently achieve satisfactory retrieval abilities, correlations between different kinds of encoded features are exploited: across the set of stored patterns, any pair of interconnected modules are simultaneously active or quiescent more often than chance, that is, adjacent modules tend to store features of correlated classes (cf. also Section \ref{statistics}). Besides, the extramodular contacts have to be more effective and/or faster in learning than the intramodular ones (cf. also Section \ref{model}). 

It is found that some spots of the network, depending on pattern to retrieve and choice of the cued fragment, cannot be reached by spreading retrieval activity (Section \ref{isles}). The volume of the modular network that is occupied by these {\it activity isles} does not depend (in large networks) on specific cue, pattern to retrieve, or retrieval dynamics, thus determining a general upper-bound to the retrieval quality. The oscillatory mechanism nearly saturates this optimal bound (cf. also Section \ref{dynamics}).

The simpler case in which local features are in one-to-one correspondence with global patterns has been also discussed (Section \ref{isles}). The retrieval dynamics in this case is simpler, but the putative `cognitive' advantage of reusing stored features for several patterns is lost, of course. In fact, during the oscillatory stage the retrieval quality of the model with feature-sharing is not too far below (in the `artificial' case, almost equal to) the best value of the model without feature-sharing.

The assumptions on the architecture of the connections and on the neurophysiological parameters used in the construction of the model seem in accord with available experimental data. The constraints necessary to reproducing proper memory retrieval also seem compatible with biological evidence. In particular, the request for active modules to be almost as noisy as the quiescent ones determines the neuronal activity sparseness $a=0.027$, which is a value in good agreement with the present experimental literature. The model also predicts that in the real columnar networks the product $s' t_{1}$ should be found to be larger than 1 (Section \ref{dynamics}).

Certainly, a more extensive mathematical study of the retrieval dynamics would be welcome. As commented earlier in the paper, use of the techniques presently available in the field of complex systems is precluded because of some profound difficulties, that are the analogue of those found in other systems with finite connectivity of interest to physics. Hence, remarkable technical advancements may be expected to be necessary in accomplishing a complete mathematical treatment.

In the present work, a firing-rate neuron model has been used, but it may be important the adoption of a spiking neuron model. In particular, it may be of interest investigating whether the effects of the oscillatory modulation introduced in this paper could instead be obtained by appropriate timing of spiking activity.
\end{multicols}
\pagebreak

\section*{Acknowledgements}
Research at University of Li\`ege was supported by an EC Grant.

\section*{\center{Appendix A \\ 
{\normalsize{Neuronal field $h_{i_{m}}$ and memory-load noise}}}}\label{field}

Summing together the input currents of neuron $ i_{m}$, one obtains:
$$
\bea{lll}
&h_{i_{m}}& = \varphi_{m} B \tau_{m}^{1} \left(\frac{\eta_{i_{m}}^{d_{m}(1)}}{a}-1\right) 
(1-\gamma) 
b (1-a) \left[1+\sum\limits_{p\in d_{m}(1), p\neq1} \tau^{p}_{m} \right] + \\
&+& \varphi_{m} {\hat x} \left[ \left(\frac{1-a}{a}\right)^2 \alpha \tau 
(1-\gamma) 
b (1+\nu) \left\{ \left[a (B+C)^2 + (1 - a) C^2\right] \tau_{m}^{1} + (A+C)^2 (1-\tau_{m}^{1})\right\} \right]^{1/2} + \\
&+& (1-\varphi_{m}) B \tau_{m}^{p_{m}} 
\left(\frac{\eta_{i_{m}}^{d_{m}(p_{m})}}{a}-1\right) (1-\gamma) b (1-a) 
\left[ 1+\sum\limits_{p\in d_{m}(p_{m}), p\neq p_{m}} \tau^{p}_{m} \right] + \\ 
&+& (1-\varphi_{m}){\hat x}' \left[ \left(\frac{1-a}{a}\right)^2 \alpha \tau 
(1-\gamma) b (1+\nu) \left\{ \left[a (B+C)^2 + (1 -a) C^2 \right] \tau_{m}^{p_{m}} + (A+C)^2 (1-\tau_{m}^{p_{m}})\right\} \right]^{1/2} + \\ [18pt]
&+& \frac{\gamma}{\lambda \hat{k}} (1-a) B \tau_{m}^{1} 
\left(\frac{\eta_{i_{m}}^{d_{m}(1)}}{a}-1\right) \sum\limits_{n \neq m} \varphi_{n} s_{m n} \tau_{n}^{1} + \\
&+& \sum\limits_{n \neq m} \varphi_{n} s_{m n} {\hat y}_{n} \left[ 
\frac{\alpha \gamma \tau t_{1}}{\lambda^2 \hat{k}}\left(\frac{1-a}{a}\right)^2 \left\{ \left[ a 
(B+C)^2 + (1 - a) C^2 \right] \tau_{n}^{1} + (A+C)^2 (1-\tau_{n}^{1})\right\} \right]^{1/2} + \\
&+& \frac{\gamma}{\lambda \hat{k}} (1-a) B \sum\limits_{n \neq m} \varphi_{n} s_{m n} \tau_{n}^{1} \sum\limits_{p\in d_{n}(1), p\neq1} \tau^{p}_{m} \tau^{p}_{n} \left(\frac{\eta_{i_{m}}^{d_{m}(p)}}{a}-1\right) + \\
&+& \frac{\gamma}{\lambda \hat{k}} (1-a) B \sum\limits_{n \neq m} (1-\varphi_{n}) s_{m n} \tau_{n}^{p_{n}} \sum\limits_{p\in d_{n}(p_{n})} \tau^{p}_{m} \tau^{p}_{n} \left(\frac{\eta_{i_{m}}^{d_{m}(p)}}{a}-1\right) + \\
&+& \sum\limits_{n \neq m} (1-\varphi_{n}) s_{m n} {\hat z}_{n} \left[ 
\frac{\alpha \gamma \tau t_{1}}{\lambda^2 \hat{k}} \left(\frac{1-a}{a}\right)^2 \left\{ \left[ a (B+C)^2 + (1 - a) C^2 \right] \tau_{n}^{p_{n}} + (A+C)^2 (1-\tau_{n}^{p_{n}})\right\} \right]^{1/2},
\eea
$$
where $p_{m}$ indicates the pattern whose feature is retrieved in module $m$ (so $p_{m}=1$ if module $m$ is in correct state).
 
The memory-load noise terms (containing the normal variables $\hat{x}$, $\hat{x}'$, $\hat{y}_{n}$, $\hat{z}_{n}$) can be unified in virtue of (asymptotic) independence properties, resulting in a Gaussian term with variance 
$$ 
\bea{rcl}
\sigma^{2}&=& \frac{\alpha \gamma t_{1} a \tau}{\lambda^2 \hat{k}} B^{2} 
\left(\frac{1-a}{a}\right)^{2} \left\{ b\frac{1-\gamma}{\gamma} 
\frac{\lambda^2 \hat{k}}{t_{1}} (1+\nu) \left[\frac{a (B+C)^2 + (1 - a) C^2}{a 
B^2}\tau_{m}^{p_{m}} + 
 \frac{(A+C)^{2}}{a B^{2}} (1-\tau_{m}^{p_{m}})\right] + \right. \\[.5cm]
&& \left. + \frac{a (B+C)^2 + (1 - a) C^2}{a B^2} (\hat{X}+\hat{v}) + 
\frac{(A+C)^{2}}{a B^{2}} ({\hat k} -\hat{X} -\hat{v}) \right\}, 
\eea
$$
where the abbreviations introduced in Section \ref{fieldov} (Eqs. \ref{define}) are used. 

\section*{\center{Appendix B \\
{\normalsize{Simulations}}} } \label{simulations}

Simulations were carried out on a Linux-operated PC. The first step required for the simulations is the construction of the random graph that underlies the extramodular connections; this is simply achieved by randomly drawing an edge between any pair of the $M$ modules with probability $s=s'/(M-1)$. Then, it is necessary to produce patterns of activity that obey the statistics of Eqs. \ref{stat}; to do this, use is made of the method proposed in \citeasnoun{CFMjpa}. Once these sets of quenched random variables are available, the analytical framework presented in this paper allows one to simulate the dynamics of large networks in reasonable CPU time. Because of the highly distributed nature of the memory representations and of the necessarily large size of the multi-modular network, exploring the effects of letting wrong activity survive any low-robustness time-step is presently unfeasible. In particular, it is not feasible to store a number of patterns sufficient to recreate the entire statistical distribution adopted in the model. This is the reason why the local component of the numerator of Eq. \ref{StoN1} was approximated with the constant $\nu$ in the simulations. If this approximation is not adopted, when in module $m$ a wrong feature is activated that is shared among many patterns, the low-robustness step is likely to be unable to destabilize it; in the next time-step, the wrong retrieval in module $m$ is then likely to elicit further wrong activation in its neighbours. The activity of the latter may not necessarily be stable enough to survive the next low-robustness step, and this fact may be expected to hamper further spread of wrong feature retrieval. However, studying this possibility would require the storing of the whole set of patterns, which, as mentioned, is presently unfeasible. For the same reason, it has also been assumed that, when a module is driven to wrong retrieval, it is supported by only the neighbour responsible for the drive, accordingly with the analytical model.

In the simulations of this work, $P=100$ patterns have been stored assuming $D=1$ and $\tau=0.1$, which provide $\nu\simeq10$ (cf. Section \ref{statistics}). In the framework of the dynamics introduced in Section \ref{dynamics}, this reduced distribution of features may increase the chance of robust activation of wrong features, even assuming the approximation of the local signal to a constant. For example, let $A$, $B$ and $C$ be any three modules such that modules $B$ and $C$ are both connected to module $A$, but do not connect with each other ($B$--$A$--$C$). Assume that a certain pattern $p$ activates features $a$, $b$ and $c$ respectively in modules $A$, $B$ and $C$. Also assume that another pattern $p'$ has been stored that activates features $b$ and $c$ respectively in modules $B$ and $C$, and activates feature $a'$, different from $a$, in module $A$. If, during the process of retrieval of pattern $p$, modules $B$ and $C$ retrieve respectively features $b$ and $c$, then there is significant chance that feature $a'$ is elicited in module $A$, and this activity would be robust to destabilization because its local associations had been stored previously too. In the analytical model, the probability that any pattern satisfies the requirements to play the role of pattern $p'$ is about $\tau t_{1}^{2}/D^{2}$; so, the probability that such a pattern exists in the stored set is vanishingly small (as $P, D \rightarrow \infty$). Instead, if $P=100$ and $D=1$, on average about 1.6 patterns could play the role of pattern $p'$; this may cause the retrieval of wrong features to survive the destabilization steps. In the simulations presented here, this possibility has been purposedly neglected. The reduced distribution of features and the relatively small size of $P$ do not affect significantly any of the other quantities involved in the model simulations.

\end{document}